\documentclass[aps,prb,showpacs,superscriptaddress,twocolumn]{revtex4-1}
\usepackage{graphicx}
\usepackage{amsmath}
\usepackage{amssymb}
\usepackage{bm}
\usepackage[pdftex]{hyperref}
\hypersetup{colorlinks=true,linkcolor=blue,citecolor=blue,urlcolor=blue}

\begin{document}

\title{Reply to ``Comment on `Critical Point Scaling of Ising Spin Glasses in a Magnetic Field' '' }

\author{Joonhyun Yeo}\email{jhyeo@konkuk.ac.kr}
\affiliation{Department of Physics, Konkuk University, Seoul 143-701, Korea}
\author{M. A. Moore}\email{m.a.moore@manchester.ac.uk}
\affiliation{School of Physics and Astronomy, University of Manchester,
Manchester M13 9PL, UK}
\date{\today}

\begin{abstract}
In  his  Comment, Temesv\'{a}ri  objects  to  a  remark in  our  paper
[Phys.\ Rev.\ B {\bf 91}, 104432  (2015)] that his result for the form
of the  Almeida-Thouless (AT) line  obtained in an earlier  paper with
Parisi [Nucl.\ Phys.\ B {\bf 858}, 293 (2012)] in six dimensions 
can be obtained by taking the limit
of $d \to 6$ in the equations  valid for $d>6$, but that this violated
one of the inequalities needed for their validity. He is just pointing
out that  they gave a  derivation of  the form of  the AT line  in six
dimensions    in   [Nucl.\ Phys.\ B {\bf 858}, 293 (2012)]    which    avoided    this
difficulty. However, it is still a perturbative approach, and does not
deal with  the lack of  a perturbative fixed  point found by  Bray and
Roberts [J. Phys. C {\bf 13}, 5405 (1980)] long ago.
\end{abstract}



\maketitle
 
The point which  Temesv\'{a}ri is making in  the Comment \cite{Temesv}
is not  about the main subject  of our paper \cite{YM}--  which was on
critical point  scaling -- but  on two  paragraphs in Sec.~III  of our
paper where we  described previous works on  the Almeida-Thouless (AT)
line. In  particular he is  objecting to  the statements in  our paper
concerning our Eq.~(13) for the form  of the supposed AT  line in six
dimensions.  In his  earlier  paper with  Parisi  \cite{PT}, they  had
derived the same equation. We remarked in our paper that this equation
did not  follow from the  equations valid  for $d >  6$, as to  get it
required  violating the  inequalities in  Eq. (7)  of his  Comment and
their equivalents in our own paper. He too makes the same point in his
Comment. But what he is pointing  out is that in Ref.  \onlinecite{PT}
an alternate derivation of Eq.~(13) was made, which is claimed to
give the correct  form for the assumed AT line in precisely six dimensions, which just happens to be the
  same equation that is obtained from using  the
equations for $d \to 6^+$ outside their limit of validity.

The   work   of   Ref.~\onlinecite{PT}  is   a   perturbative
renormalization group (RG)  calculation.  Bray and Roberts  \cite{BR} showed
many years ago there is no  stable fixed point for the perturbative RG
equations for the critical behavior across the supposed AT line in six
dimensions and below. They suggested  that this might imply that there
was no AT line at and below six dimensions. This was the view taken in
Ref.~\onlinecite{AT6}. What was done in Ref. \onlinecite{PT} and in the Comment\cite{Temesv} was to simply assume that there was an AT line, despite the fact that within the perturbative RG theory for $d \le 6$ it is has never proved possible to obtain it.  Until   Temesv\'{a}ri   can  overcome   this
difficulty  it  is   hard  to  take  seriously  the   claim  that  his
perturbative  calculation  of the  AT  line  for  $d  \le 6$  has  a shred of 
validity.

JY was supported by
Basic Science Research Program through the National Research Foundation 
of Korea (NRF) funded by the Ministry of Education (2014R1A1A2053362).

\end{document}